\documentclass[10pt]{article}
\usepackage{amsmath,amsfonts,amsthm,amssymb,amscd}
\usepackage[toc,page]{appendix}
\numberwithin{equation}{section}

\newcommand{\be}{\begin{equation}}
\newcommand{\ee}{\end{equation}}
\newcommand{\bs}{\begin{split}}
\newcommand{\es}{\end{split}}
\newcommand{\ba}{\begin{align}}
\newcommand{\ea}{\end{align}}
\newcommand{\basl}[1]{\begin{align}\begin{split}\label{#1}}
\newcommand{\bas}{\begin{align}\begin{split}}

\newtheorem{theo}{Theorem}[section]
\newtheorem{prop}[theo]{Proposition}

\newcommand\N{\mathbb{N}}

\newcommand\R{\mathbb{R}}
\newcommand\C{\mathbb{C}}

\title{Classical and quantum energies for interacting magnetic systems }

\author{ L. Amour, J. Nourrigat}

\date{Laboratoire de Math\'ematiques de Reims - UMR CNRS 9008  \\ \vskip 0.15cm Universit\'e de Reims Champagne-Ardenne, France}

\begin{document}

\maketitle

\begin{abstract}
\noindent The purpose of this article is to give different interpretations of the first non vanishing term (quadratic) of the ground state asymptotic expansion for a spin system in quantum electrodynamics, as the spin magnetic moments go to $0$. One of the interpretations makes a direct link with some classical physics laws. A central role is played by an operator $A_M$ acting only in the finite dimensional spin state space and making the connections with the different interpretations, and also being in close relation with the multiplicity of the ground state.
 \end{abstract}

\
\parindent=0pt
{\it Keywords:} Interacting spins, Ground state energy, Interacting magnets, Classical energy,  Higher spins, Quantum electrodynamics, Quantum radiative corrections, Multiplicity of the ground states. 

\


\parindent=0pt
\parindent = 0 cm

\parskip 10pt
\baselineskip 12pt

\section{ Introduction.}\label{s1}

The spin of one of several static interacting particles (fixed atomic nuclei) can be described in the quantum electrodynamic  (QED)  setting by a Hamiltonian operator given in \cite{CTD01} (see also \cite{Reu}). Besides, it is common to represent these static $\frac{1}{2}-$spins as magnets able to turn around fixed points and interacting according to the classical physics laws. It our aim in this paper to make a link between these two points of view by studying the ground state energy of the QED Hamiltonian, recalled in Section \ref{s2} and having a non degenerate eigenvalue  as  the infimum of the spectrum (in the case of several spins). Each spin particle in this Hamiltonian has a magnetic moment  playing  the role here of a coupling constant in the interaction of the particle with the quantized electromagnetic field. There is no external magnetic in the model under consideration in the purpose to fully  focus on the interaction between  spins themselves. We recall that, in the case of non zero external magnetic fields, the ground state energy is an analytic function of the coupling constant (see \cite{HH11,Ara14}).

The main objective of this work is to provide different expressions of the first non vanishing term (which is of order two) of the asymptotic expansion of the  QED model  ground state energy for small magnetic moments. One of these expressions makes a link with classical magnetostatics.

Let us introduce some notations in order to be more specific. The number of spin particles is denoted by $P$ which is fixed throughout this work, $M^{[\lambda]}$, $\lambda=1,\dots,P$,  stands for the magnetic moment of the $\lambda$-th particle, we write $M=(M^{[1]},\dots,M^{[P]})$ and  the   QED Hamiltonian is denoted by $H(M)$. The Hilbert spaces of spin states and  of free photon states are respectively ${\cal H}_{\rm sp} = (\C^2) ^{\otimes P}$ and  ${\cal H}_{\rm ph}$ which is a Fock space (see Section \ref{s2}). The Hamiltonian $H(M)$ is  an unbounded operator in ${\cal H}_{\rm tot} = {\cal H}_{\rm ph} \otimes {\cal H}_{\rm sp}$ and its spectrum is denoted by $\sigma (H(M))$.

We shall prove  that there exists a quadratic  form  $M \mapsto E_2 (M)$ 
   on  $\R^P$ satisfying the identity $ \inf \sigma (H(M))  = E_2(M) + {\cal O} (|M|^3 )$ for small $|M|$, and our goal in this article is to provide distinct expressions of  $E_2 (M)$ interpreted in different ways.  Note that $E_2(M)$ can be considered as the first radiative correction of the ground state energy which is zero when $M$ vanishes.

      {\it 1.} First, we prove that  $E_2(M)$ is the smallest eigenvalue
      of some operator  $A_M$ acting in the finite dimensional spin space ${\cal H}_{\rm sp}$. 
      The operator $A_M$ is defined in  (\ref{def-A-M}), quadratic of  
      $M=(M^{[1]},\dots,M^{[P]})$ and has an important role here.
      We prove (Theorem \ref{prop-2}) that: 
\be\label{hess}  \inf \sigma ( H(M)) = \inf \sigma  ( A_M) + {\cal O} (|M|^3) \ee
as $|M|$ goes to $0$ where
 $ |M| = \big (\sum _{\lambda = 1 } ^N   |M^{[\lambda]} |^2 \big )^{\frac{1}{2}}$.

      {\it 2.} The quadratic form $X\mapsto < A_M X , X>$   defined for any $X$ in the spin space ${\cal H}_{\rm sp}$ 
   is non positive. This quadratic form has  several interpretations. First, for each fixed  $X\in {\cal H}_{\rm sp}$ and $\rho>0$, we consider the ball $B_{\rho |M|}  (X)$ 
    in the domain of $H(M)$ (with the norm defined in (\ref{D-H})) centered at  $\Psi_0 \otimes X$  with radius $ \rho  |M| |X|$ where $\Psi_0$ denotes the vacuum state in the photon state space  $ {\cal H}_{\rm ph}$.
     We prove  (Proposition \ref{prop-1}) for sufficiently large $\rho $ that, the quadratic form $M \mapsto < A_M X , X >$ is the first non vanishing term (of order $2$) of the asymptotic expansion for small $|M|$ of: 
 \be\label{borne-inf}  M \mapsto \inf _{f \in  B_{\rho |M|}  (X) } < H(M) f, f>.  \ee 
    
     {\it 3.} In another direction,  $< A_M X , X>$ is expressed using Maxwell equations, but taking values in ${\cal H}_{\rm sp}^3$ instead of usually $\R^3$, for every fixed $X\in {\cal H}_{\rm sp}$.
More precisely,  a current density $x \mapsto {\bf j}^{\rm \bf vect} (x , X)$  taking values into ${\cal H}_{\rm sp}^3$ is defined for any $X\in {\cal H}_{\rm sp}$ giving a magnetic field  
$x \mapsto {\bf B}^{\rm \bf vect} (x , X)$  taking values into ${\cal H}_{\rm sp}^3$ (also called vector-valued here).  The point is that (Theorem \ref{th-egal}),  for any $X\in {\cal H}_{\rm sp}$, the energy of this vector-valued magnetic field, defined by analogy with the classical formula, is simply the opposite of $<A_M X , X>$. 
  
    {\it 4.} It is however for specific sates  $X\in {\cal H}_{\rm sp}$, namely tensor products
 $X = V^{[1]} \otimes \cdots \otimes V^{[P]}$
  with the $V^{[\lambda]}$ in $\C^2$  and normalized, also called product states  or disentangled states, that the function
   $< A_M X , X>$ displays a  very common form.  To this end, first recall 
 that, with any product state $X\in  {\cal H}_{\rm sp}$, the Hopf fibration defines $S(X) = ( S^{[1]} (X), \dots, S^{[P]} (X) )$, a set of $P$ elements  belonging to the unit sphere $S_2$ of $\R^3$. Second, with any set $S = ( S^{[1]} , \dots, S^{[P]}  )\in(S_2)^P$, the classical current density ${\bf j} ^{\rm \bf class} (x , S)$ is defined by  (\ref{tenso-3}), for any magnetic moments $M^{[\lambda]}$ ($\lambda=1,\dots,P$). Let us mention at this step that,  the ultraviolet cut-off of the QED Hamiltonian $H(M)$ appears through its inverse Fourier transform in the definition of the classical current density  (\ref{tenso-3}). Thus, ${\bf j} ^{\rm \bf class} (x , S(X))$ can be viewed as the  current density of a system of magnets corresponding to the initial spin system where  $M^{[\lambda]}>0$ and   $S^{[\lambda]}(X)$  respectively are  the magnetic moment and the orientation of the $\lambda$-th magnet. The point is now that, for every product state $X$, we prove (Theorem \ref{class-deug}) the validity of a simple relation between $< A_M X , X> $ and the magnetic energy defined with classical physics laws associated coming from the classical current density ${\bf j} ^{\rm \bf class} (x , S(X))$. Thus, in some sense, this shows that the classical physics laws on a system of interacting magnets  at equilibrium are derived from QED. This also shows that the QED model in  \cite{CTD01} indeed encompasses the interaction between particles, without including ad hoc additional interaction terms.

 Finally and still regarding the operator $A_M$, we prove (Theorem  \ref{theo-4}) that $A_M$ plays a significant role in the study of the multiplicity of the ground states of $H(M)$. 
 
Points $1$-$4$  are developed below. The main results of this article are thus Theorem  \ref{prop-2} together with Theorems \ref{th-egal}-\ref{class-deug} and Theorem  \ref{theo-4} and are precisely stated in Section \ref{s2}. Additionally, the case of higher spin particles is investigated at the end of Section \ref{s2}.

 As complementary results concerning the QED model of \cite{CTD01}, note that regarding the issue of time dynamics evolution, different approximations in \cite{A-J-N-2,A-N} leads to differential systems simpler than the initial model, but without connection to standard electromagnetism, and see \cite{A-J-N-3}  concerning the  localization in position of the ground state photons.

  \section{Statements of the results.}\label{s2}  
  
  We use the notations $|\cdot|$ and $\Vert\cdot\Vert$  for the all the norms respectively in the case of finite and infinite dimension since they always refer to the canonical norm of the space under consideration. We also use $<\cdot , \cdot>$ for all scalar products except for $\R^3$ where we use the notation $\cdot\,$ which is also used when one of the element is in $\R^3$ and the other in ${\cal H}_{\rm sp}^3$, and similarly for the cross product $\times$ used with two elements of $\R^3$ or one element in $\R^3$ and the other in ${\cal H}_{\rm sp}^3$. Still for notations, we use $C$ as a constant that may change from line to line but not depending on varying parameters such as $M$ and $X$.
   
   We first recall the definition of $H(M)$.
   
  {\it  Definition of the Hilbert space.}
  
    The state Hilbert space ${\cal H}_{\rm ph}$ of non interacting photons
   is the symetrized  Fock space over the single photon state space $\mathfrak{H}$,  that is, 
    ${\cal H}_{\rm ph} = {\cal F}_{s}( \mathfrak{H} )$ also often denoted 
     $\Gamma (\mathfrak{H} )$ where (\cite{LL04}):
   $$  \mathfrak{H} = \{  V \in L^2 ( \R^3 ,  \R^3 ),\quad 
   k\cdot V(k) = 0 \ \text{for a.e. } k\in\R^3 \}. $$
   In ${\cal H}_{\rm ph}$, we use operators ${\rm d}\Gamma (T)$ for bounded or unbounded operators $T$ in $\mathfrak{H}$ and operators 
   $\Phi_{\rm S} (V)$ (Segal fields) for $V\in \mathfrak{H}$  (\cite{RSII}). The photon vacuum state in ${\cal H}_{\rm ph}$ is denoted here by  $\Psi_0$.
  
   The Hilbert space for a system of $P$ static $\frac{1}{2}$-spin particles in the presence of photons is the completed tensor product space
    ${\cal H}_{\rm tot} = {\cal H}_{\rm ph} \otimes {\cal H}_{\rm sp}$ where
    ${\cal H}_{\rm sp} = ({\bf C}^2 )^{\otimes P}$.  
    
Recall the  Pauli matrices: 
    \be\label{Pauli} \sigma_1 = \begin{pmatrix}  0 & 1 \\ 1 & 0   \end{pmatrix},
\quad
\sigma_2 = \begin{pmatrix} 0 & -i \\ i & 0   \end{pmatrix},
\quad
\sigma_3 = \begin{pmatrix}  1 & 0 \\ 0 & -1  \end{pmatrix},\ee
and set:
 \be\label{H-M-4}  \sigma_m ^{[\lambda]}  = I \otimes \cdots \otimes I\otimes \sigma _m\otimes I \otimes \cdots I,\quad
m=1,2,3 ,\quad  \lambda =1,\dots,P, \ee
where $\sigma _m$ is located at the $\lambda ^{th}$ position.

   {\it  Definition of the Hamiltonian.}
   
      The system under consideration constituted with $P$ static $\frac{1}{2}$-spin particles interacting with the quantized electromagnetic field is described by an Hamiltonian operator  
    $ H(M^{[1]}, \dots, M^{[P ]})$ in ${\cal H}_{\rm tot}$ where the $M^{[\lambda]}$
  are the magnetic moments of the particles (\cite{CTD01}). This Hamiltonian  is a selfadjoint extension of the following operator initially defined on a dense subset:
  \be\label{H-M-1}  H(M) = H_{\rm ph} \otimes I +  H_{\rm int}(M),\qquad  M= (M^{[1]} , \dots, M^{[P]} ), \ee 
 where:
   \be\label{H-M-2} H_{\rm ph} =  \  {\rm d}\Gamma (M_{\omega }), \ee 
    with $M_{\omega }$ being the multiplication operator  by the function $\omega (k)= |k|$ in $\mathfrak{H}$ and where:
   \be\label{H-M-3}  H_{\rm int}(M) =  \sum _{\lambda = 1 }^P \sum _{m= 1 }^3 M^{[\lambda]} \Phi_{\rm S} ( B_{m , x^{[\lambda]} })
   \otimes \sigma_m ^{[\lambda]},  \ee 
     with   $x^{[\lambda]}\in \R^3$  being the position of $\lambda$-th particle and where $B_{m,x}\in \mathfrak{H}$ is given by:
\be\label{Bjx}   B_{m,x}(k) = {i  \phi(|k| )|k|^{1\over 2} \over (2\pi)^{3\over 2}}
e^{-i(   k\cdot x   )} {k\times e_m \over |k|},\quad m=1,2,3,\quad x\in\R^3,\quad  k\in\R^3\backslash\{0\}, \ee
where $(e_1,e_2,e_3)$ is the canonical basis of $\R^3$.
   
The above function $\phi $ (ultraviolet cutoff) is assumed to belongs to the Schwarz space ${\cal S} (\R)$ and supposed to be a radial function.
All universal physics constants are taken equal to $1$ since they play no role in the following. Note that it is not necessary for this Hamiltonian to add an extra term representing the interaction between spins since this interaction is realized exclusively through the  spin-photon interaction $H_{\rm int}(M)$. 

We recall that the definition of $H(M)$ as a selfadjoint operator   is standard and follows the next inequality for any $B$ when  $B$ and $M_{\omega }^{-1/2} B$ belong to $\mathfrak{H}$:
 \be\label{BFS}  \Vert \Phi_{\rm S}(B) f \Vert   \leq  2^{\frac{1}{2}}
 \Vert  M_{\omega }^{-1/2} B \Vert \  \Vert  H_{\rm ph}^{1/2} f \Vert  + 2^{-\frac{1}{2}}
  \Vert B \Vert \  \Vert f \Vert,\quad f\in D(H_{\rm ph}) \ee
(see, $e.g$, (1.8)(1.9) in \cite{F-H-02}).
Then,   Kato Rellich Theorem shows that    $H(M)$ initially defined on  a dense domain 
by (\ref{H-M-1})-(\ref{Bjx}) is essentially selfadjoint and the domain of its unique selfadjoint extension is  $D( H_{\rm ph} \otimes I)$.
Thus $D(H(M))=D( H_{\rm ph} \otimes I)$ is independent of $M$ and is therefore from now on denoted by $D(H)$. The norm of $D(H)$ is involved in the sequel and is  given by:
\be\label{D-H}  \Vert f \Vert _{D(H)}^2  =  \Vert ( H_{\rm ph} \otimes I) f \Vert ^2 +  \Vert f \Vert ^2. \ee 
Regarding selfadjointness issues, see \cite{C-G} and also \cite{F-H-02}. For the existence of ground states, see \cite{C-G} (see also \cite{Gross-1,S-1989,AH97}).

{\it Equivalent definitions of the operator $A_M$ and of the quadratic form 
$X\mapsto < A_M X , X >$.} 

As already mentioned, we shall define an operator 
 $A_M$ acting in a finite dimensional space, namely in ${\cal H}_{\rm sp}$, quadratic in 
 $M= (M^{[1]} , \dots, M^{[P]} )$, with the property that  its smallest eigenvalue
 is also the first non vanishing term (quadratic) of the asymptotic expansion for small $|M|$ of the ground state energy  of $H(M)$, seen as a function of $M$. 
 
 We shall now be more specific on the Points {1-4} of Section \ref{s1}

{\it 1.}   Let  $E: {\cal H}_{\rm sp} \rightarrow {\cal H}_{\rm tot}$ 
be the mapping defined by $EX = \Psi_0 \otimes X$ for all $X\in {\cal H}_{\rm sp}$ and denote its adjoint by $E^{\star}: {\cal H}_{\rm tot} \rightarrow {\cal H}_{\rm sp}$. We set:
\be\label{def-A-M}  A_M  X = -  E^{\star}  H_{\rm int} ( ({\rm d}\Gamma ( M_{\omega}) )^{-1} \otimes I )
 \ H_{\rm int} E X,\quad X\in {\cal H}_{\rm sp}.\ee 
Proposition \ref{expl-AM} shows that this definition makes sense and makes explicit this operator. It is also clear that $<A_MX , X > \leq 0$ for any $X\in {\cal H}_{\rm sp}$. 

We now report three interpretations of  $< A_M X , X >$ for every fixed $X\in {\cal H}_{\rm sp}$.

{\it 2.} The first interpretation of  $< A_M X , X >$  is to be equal to the infimum of the QED energy on a specific ball in the domain of $H(M)$. To this end, 
 set for any  $X\in {\cal H}_{\rm sp}$ and every $\rho >0$:
 $$ B_{\rho}  (X) = \{ f\in D(H), \quad \Vert f - \Psi_0  \otimes X \Vert _{D(H)}
 \leq  \rho  \ |X|   \}.  $$
One checks that there exists $K>0$ satisfying for all $X\in {\cal H}_{\rm sp}$:
  \be\label{def-K} \Vert  (  {\rm d}\Gamma (M_{\omega}) )^{-1}\otimes I )   H_{\rm int}  ( \Psi_0 \otimes X) \Vert_{D(H)}
 \leq K \ |M| \ |X|. \ee
With these notations, we have (see the proof in Section \ref{s3}):
 \begin{prop}\label{prop-1} There exists $C>0$ such that, if $\rho \geq K$ 
and if $|M| < 1/\rho  $ then:
 $$\Big  | < A_M X  , X > - \inf _{f \in  B_{\rho |M|}  (X) } < H(M) f, f>    
 \Big | \leq C \rho ^2  M^3 |X|^2. $$
That is to say, for $\rho\geq K$, $ < A_M X , X > $ is the first non vanishing term (quadratic) of the  asymptotic expansion for small $|M|$ of the function (\ref{borne-inf}). In that case,  
this term is independant of  $\rho$. 
 \end{prop}

{\it 3.} The second interpretation of  $< A_M X , X>$ calls upon Maxwell equations but for a current density and  a magnetic field taking values in ${\cal H}_{\rm sp}^3$. We can call them throughout the article respectively, vector-valued Maxwell equations, vector-valued current density and vector-valued magnetic field. A vector-valued current density $x \mapsto {\bf j}^{\rm \bf vect} (x , X)$ is associated with any $X\in {\cal H}_{\rm sp}$.
Precisely, we define ${\bf j}^{\rm \bf vect} $ as a mapping from $\R^3  \times {\cal H}_{\rm sp}$ 
into ${\cal H}_{\rm sp}^3$  by:
\be\label{dens-cour}  {\bf j}^{\rm \bf vect}  (x , X) = \sum _{\lambda = 1}^P M ^{[\lambda]} \nabla \rho ( x - x^{[\lambda]} )\times {\bf \sigma}  ^{[\lambda]}  X,\quad X\in {\cal H}_{\rm sp}\ee 
where:
\be\label{rho-phi}  \rho (x) = (2\pi)^{-3} \int _{\R^3} \phi (|k|) e^{i k\cdot x}  dk. \ee
 We remind that $\phi$ is the smooth ultraviolet cut-off in  (\ref{Bjx}).
We above use the notation $   {\bf \sigma}  ^{[\lambda]}  X = (  \sigma _1 ^{[\lambda]}  X  , 
 \sigma _2 ^{[\lambda]}  X ,  \sigma _3 ^{[\lambda]}  X  )$ which is an element of  ${\cal H}_{\rm sp}^3$. 
 Then, this  vector-valued current density defines a 
vector-valued vector field
 ${\bf B}^{\rm \bf vect} (x , X)$  through the  vector-valued Maxwell equations:
 $$ {\nabla}\cdot {\bf B^{\rm \bf vect} } (x, X) = 0,\qquad {\nabla}\times {\bf B^{\rm \bf vect}} (x, X) = {\bf j}^{\rm \bf vect} (x, X).$$
We shall prove the property below.
 \begin{theo}\label{th-egal} For every $X\in {\cal H}_{\rm sp}$, the following identity holds true:
 \be\label{maxwell-vect}  < A_M X , X > = -  \frac {1}  {2} \int _{\R^3 }|  {\bf B^{\rm \bf vect} } (x, X)|^2 dx.   \ee 
   \end{theo} 
 The norm in the above right hand side is the  ${\cal H}_{\rm sp}^3$ norm.

{\it 4.} The third link is with  classical physics.  We now make explicit $ <A_M X , X>$ in the special case  where 
 $X$ is a tensor product (product state): 
 \be\label{tenso}  X = V^{[1]} \otimes \cdots \otimes V^{[P]}, \qquad V^{[\lambda]} \in \C^2,\qquad
 |V^{[\lambda]}| = 1,\qquad \lambda=1,\dots,P. \ee 
 With any product state $X$, we associate  the following $P$ elements $S^{ [\lambda]}(X) $ of the unit sphere $S_2$  of 
 $\R^3$:
 \be\label{tenso-2} S^{ [\lambda]} (X)  = ( < \sigma _1 V^{[\lambda]} , V^{[\lambda]} > ,  < \sigma _2 V^{[\lambda]} , V^{[\lambda]} > ,  < \sigma _3 V_V^{[\lambda]}  , V^{[\lambda]} > ),\qquad \lambda=1,\dots,P. \ee
 This is Hopf fibration. Then, for any  $S = (S^{ [1]} , \dots, S ^{ [P]})\in (S_2)^P$, 
 one defines the classical current density (taking values in $\R^3$):  
\be\label{tenso-3}   {\bf j} ^{\rm \bf class} (x , S) = \sum _{\lambda = 1}^P M ^{[\lambda]} \nabla \rho ( x - x^{[\lambda]} )
\times S^{[\lambda] }   \ee 
where $\rho$ is related to the  ultraviolet cut-off by (\ref{rho-phi}). 
Next, $ {\bf B} ^{\rm \bf class} (x , S)$ denotes the vector fields given by the classical Maxwell equations:
 $$ {\nabla}\cdot {\bf B^{\rm \bf class} } (x, S) = 0, \qquad {\nabla}\times {\bf B^{\rm \bf class}} (x, S) =
  {\bf j}^{\rm \bf class}  (x, S).$$
We these notations, we have the following result.

 \begin{theo}\label{class-deug}  Let $X$ be defined in (\ref{tenso}) and define $S(X)$ 
by (\ref{tenso-2}). Then, 
\be\label{factor1} < A_M X , X >  =  -  \frac {1}  {2} \int _{\R^3 }|  {\bf B^{\rm \bf class} } (x, S(X))|^2 dx  - A_{11} (0) \sum _{\lambda=1}^P (M^{[\lambda]})^2,  \ee
 where the real  number $A_{11} (0)$ is given by (\ref{A-j-m}). 
  \end{theo} 
  
 Theorems  \ref{th-egal} and \ref{class-deug} are proved in Section \ref{s3}.
 
 Knowing these three interpretations of $ <A_M X , X>$, we now turn to the statement saying that the infimum of $A_M$ acting on the spin space is actually the first non vanishing term (quadratic) of the asymptotic expansion of the ground state energy of the QED Hamiltonian, as the magnetic moments go to zero.

 {\it Asymptotic expansion of the ground state energy.}  
 
We have the following identity:

\begin{theo}\label{prop-2}  If $A_M$ is defined in (\ref{def-A-M}) then equality  (\ref{hess}) holds true as   $|M|$ tends to $0$.
\end{theo} 
This result is proved in Section \ref{s4}.

{\it Multiplicity of the ground state.} 

We now assume, only in this part of the article,  that the magnetic moment coupling constants  are all equal, that is, $M^{[1]} = \dots = M^{[P]}$ which is denoted by $g$. Thus, the operator $A_M=g^2 A_1$ where $A_1$ is a selfadjoint operator in ${\cal H}_{\rm sp}$.

\begin{theo}\label{theo-4}  We have the two properties:

$(i)$ If there exists a sequence of normalized ground states $U(M_n)$ converging to a limit for some  sequence $(M_n)$ converging to $0$ (with $M^{[1]}_n = \dots = M^{[P]}_n$)   then this limit is of the form  $\Psi_0 \otimes X$ where $X$ is an eigenvector of $A_1$ corresponding to the smallest eigenvalue of $A_1$.

$(ii)$ The multiplicity of the ground state of $H(M)$ is smaller than or equal to the multiplicity of the smallest eigenvalue of $A_1$, for all small enough $M$ with $M^{[1]} = \dots = M^{[P]}$.
 \end{theo}
Note that  in the case $P=1$ of a single spin, the hypothesis of $A_1$ in  point $(ii)$ is not satisfied. Indeed,  in that case the multiplicity of the ground state is greater than or equal two, according to  \cite{L-M-S}. See also \cite{AH97,Hir05} for the multiplicity of ground states issue.

{\it Higher spin particles.} 

In the remaining part of this section, we observe that the case of higher spin particles can be considered similarly to the case of spin $\frac{1}{2}$ particles and we underline here the main modifications.
 We replace  $\C^2$
by $V_s = \C ^{2s + 1}$ for particles with  half-integer spin $s = \frac{1}{2}, 1 , \frac{3}{2},\dots$. We denote by   $\rho_s$  the standard irreducible representation in $V_s$ of the group
 $ G =SU(2)$  and by  ${\cal G}$ its Lie algebra. The set $(u_1 , u_2 , u_3)$ stands for a basis of
 ${\cal G}$ satisfying the commutation relations $[u_1 , u_2 ] = 2  u_3$ (with cyclic permutations on the indices). The Pauli matrices are then replaced by the selfadjoint matrices  
 $\sigma _j (s) = id\rho_s(u_j)$, for $j=1,2,3$. We recall that the Casimir operator satisfies: 
 \be\label{casimir} \sum _{j= 1}^3 \sigma _j(s) ^2 = 4 s (s+1) I.\ee 
The spin state space of  $P$ spin $s$ particles  becomes ${\cal H}_{\rm sp} = (V_s)^{\otimes P}$. For any $j=1,2,3$ and $\lambda=1,\dots, P$,  
   $ \sigma_j(s) ^{[\lambda]} $ is defined as in (\ref{H-M-4}) and the remaining parts of the definitions of  $H_{\rm int}$ and $H(M)$ are identical  to the case $s=\frac{1}{2}$. The definition of the operator    $A_M$ is unchanged. In that context of higher spins, the most interesting point is the modification of the constant involved in  Theorem \ref{class-deug}, that is the modification of the factor $1$ in front of $A_{11} (0) \sum _{\lambda=1}^P (M^{[\lambda]})^2$ in (\ref{factor1}). To this end, we have to consider product states of the form: 
    \be\label{tenso-s}  X = V^{ [1]} \otimes \cdots \otimes V^{ [P]},\quad V^{ [\lambda]} \in \Omega,\quad \lambda=1,\dots, P,  \ee
    for some $\Omega \subset V_s$ that we now define.
  We fix  $X_0\in V_s$ satisfying
   $\sum _{j=1}^3 < \sigma_j(s) X_0 , X_0 > ^2 = 1 $
and set:  
\be\label{Omega-def}
\Omega =\big\{ X\in V_s,\quad X=\rho_s(g) X_0,\quad g\in G\big \}.
\ee
The Hopf mapping for spin $s$ on $\Omega$
is the application $\pi : \Omega \rightarrow \R^3$
defined for all $X\in \Omega$ by:
\be\label{Hopf-s} \pi (X) = \big ( < \sigma _1(s) X , X > , < \sigma _2(s) X , X >  , < \sigma _3(s) X , X > \big ). \ee
It is  probably very well-known that $\pi$ maps  $\Omega$ into the unit sphere $S_2$ but we give a proof for the reader convenience at the end of Section \ref{s3} (Theorem \ref{Hopf-S2}). Next, with any $X$ expressed as in (\ref{tenso-s}), we define $S(X)$  the set of $P$ elements of $S_2$ given by:
 \be\label{tenso-2-s} S(X) = ( S^{ [1]} (X) , \dots, S^{ [P]} (X)),\quad S^{ [\lambda]} (X)  = \pi (V^{ [\lambda]}),\ \lambda=1,\dots,P. \ee
Then, in the case of half-integer spin particles, one still have a simple interpretation of the quadratic form 
 $<A_M X , X>$ for $X$ as in (\ref{tenso-s}) with the next result extending Theorem \ref{class-deug}.
  \begin{theo}\label{class-deug-s} 
   Fix any half-integer $s$. Let $X$ be under the form (\ref{tenso-s}) with $\Omega$ defined in (\ref{Omega-def}) and  let $S(X)$  be given by (\ref{Hopf-s})(\ref{tenso-2-s}). Then, the following identity holds true: 
  $$ < A_M X , X >  =  -  \frac {1}  {2} \int _{\R^3 }|  {\bf B^{\rm \bf class} } (x, S(X))|^2 dx  - C(s)A_{11} (0) \sum _{\lambda=1}^P (M^{[\lambda]})^2,  $$
 where: $$C(s)=2s(s+1)-\frac{1}{2}.$$ 
  \end{theo} 
The proof of Theorem \ref{class-deug-s} is a straightforward modification of the proof of Theorem \ref{class-deug} using (\ref{casimir}) and is therefore omitted.

In Section \ref{s3}, Proposition \ref{prop-1} and Theorems \ref{th-egal}-\ref{class-deug} are proved. We derive Theorem \ref{prop-2} in Section \ref{s4}. The proof of Theorem  \ref{theo-4} is completed in Section \ref{s5}.

\section{Connections between the four definitions of the energy.}\label{s3}

We denote
by  $\Pi_0f$   the orthogonal projection of $f $ on the subspace $\Psi_0 
\otimes {\cal H}_{\rm sp} $ for any  $f\in {\cal H}_{\rm tot} = {\cal H}_{\rm ph}  \otimes {\cal H}_{\rm sp} $ and we set  $\Pi _{\perp } = I - \Pi_0$. Thus, 
$\Pi_0f$  is of the form $\Psi_0 \otimes X$ and we denote by $\Pi_{\rm sp} f $ the element $X\in {\cal H}_{\rm sp} $.

{\it Proof of Proposition \ref{prop-1}. } $(i)$
For any  $X\in {\cal H}_{\rm sp}$ and each $M\in \R^P$, we set:
 $$ \varphi (M , X) = \Psi_0 \otimes X -  u_M(X), $$ where:
   $$ u_M(X) = ( ( {\rm d}\Gamma (M_{\omega}) )^{-1}\otimes I )   H_{\rm int}  ( \Psi_0 \otimes X). $$

The element $ \varphi (M , X)$ belongs to $B_{K|M|}  (X)$ where $K$ is defined in  (\ref{def-K}). One checks that:
 $$ < H(M) \varphi (M  , X) , \varphi (M  , X) > =    < A_M X , X> + 
 < H_{\rm int}  u_M(X) ,  u_M(X) >. $$
Besides, $ < H_{\rm int}  u_M(X) ,  u_M(X) >  = 0$ since the two elements $u_M(X)$ and $ H_{\rm int}  u_M(X)$ belong  to two orthogonal sectors of the Fock space decomposition. Consequently, if $\rho \geq K$:
 $$\inf  _{ f \in B_{\rho |M| } (X) } < H(M) f , f >   \leq \   < A_M X  , X >. $$
 $(ii)$ We verify that:
 $$  < A_M X , X> = \inf_{f \in  D( H)} \ < ( H_{\rm ph}\otimes I) f , f > +
 2 {\rm Re} < H_{\rm int } (\Psi_0 \otimes X), f >. $$
 Thus, for any $f\in D(H)$:
 $$ < A_M \Pi_{\rm sp} f , \Pi_{\rm sp} f > \leq
 ( H_{\rm ph}\otimes I) f , f > +
 2 {\rm Re} < H_{\rm int } \Pi_0 f, f >.$$
Besides:
$$ < H(M) f , f > =  < ( H_{\rm ph}\otimes I) f , f > +
 2 {\rm Re} < H_{\rm int } (\Pi_0 f), f >  + < H_{\rm int} \Pi_{\perp} f , \Pi_{\perp} f >. $$
Thus:
 $$ < A_M \Pi_{\rm sp} f , \Pi_{\rm sp} f > \leq < H(M) f , f >  +
 |  < H_{\rm int} \Pi_{\perp} f , \Pi_{\perp} f > |. $$
According to (\ref{BFS}):
 $$  \Vert H_{\rm int} g \Vert  \leq  C |M | ( \Vert ( H_{\rm ph} \otimes I) g \Vert  +
  \Vert  g \Vert ).  $$
If $f$ belongs to $B_{\rho |M|}  (X)$ then:
$$ | \Pi_{\rm sp} f - X | \leq \rho |M| | X | ,\quad \Vert \Pi_{\perp} f  \Vert \leq
\rho |M| | X |  ,\quad \Vert ( H_{\rm ph} \otimes  I )  \Pi_{\perp} f  \Vert \leq
 \rho |M| | X |.$$
As a consequence, for every $f$ belonging to $B_{\rho |M|}  (X)$: 
 $$ |  < H_{\rm int} \Pi_{\perp} f , \Pi_{\perp} f > |  \leq C  \rho^2  |M|^3 | X |^2.   $$
Similarly,   if $f\in B_{\rho |M|}  (X)$ and if $ \rho |  M | \leq 1$ then:
\begin{align*}
< A_M X , X>\   &\leq\  < A_M \Pi_{\rm sp} f , \Pi_{\rm sp} f >  + C |M| ^2 (
 | \Pi_{\rm sp} f - X |^2 + | \Pi_{\rm sp} f - X | \ |  X | ) \\
   &\leq \ < A_M \Pi_{\rm sp} f , \Pi_{\rm sp} f >  +C  \rho  |  M |^3  |  X |^2.
   \end{align*}
 
Thus:
 $$ < A_M X, X>\  \leq  \  < H(M) f , f > + C \rho ^2 |M | ^3 |X | ^2, $$
which proves Proposition \ref{prop-1}.\hfill $\Box$

The following Proposition is used to 
give an explicit expression of $A_M$ 
defined in (\ref{def-A-M})  and to prove that this definition makes sense. It will be also useful for the proof of 
Theorem \ref{th-egal}.

\begin{prop}\label{expl-AM} We have:
 $$ A_M  X  = - \frac {1}  {2} \sum _{\lambda , \mu \leq P }  \sum _{j, m \leq 3  }  M^{[\lambda]}   M^{[\mu]}
  A_{jm} (x^{[\mu]}  - x^{[\lambda]}  ) \sigma _m ^{[\mu]}    \sigma _j ^{[\lambda]}  X    $$
 where:
\be\label{A-j-m}   A_{jm} (x ) = ( 2 \pi )^{-3} \int _{\R^3}  | \phi(|k|)|^2  e^{-i   k\cdot x }
  \frac { \delta _{jm} |k|^2 - k_j k_m  } {  |k|^2 }  dk.  \ee

\end{prop}
The elements $A_{jm}(x)$ are related to the standard transverse delta function \cite{F} (smeared out with $ | \phi|^2$).

{\it Proof.}  We see that:
$$ H_{\rm int}  ( \Psi_0 \otimes X) = \frac  {1} {\sqrt 2 }  \sum  _{\lambda  \leq P  }  \sum _{j \leq 3  }
 M^{[\lambda]} B_{j , x^{[\lambda]} }  \otimes \sigma _j ^{[\lambda]}  X. $$
Thus, with $A_M$  defined in (\ref{def-A-M}) and for every  $X$ and $Y$ 
in ${\cal H}_{\rm sp}$:
 $$  <  A_M  X , Y >  = - \frac  {1} {2 }   \sum _{\lambda , \mu \leq P  }  \sum _{j, m \leq 3  }  M^{[\lambda]}   M^{[\mu]}
< ({\rm d}\Gamma ( M_{\omega}) )^{-1} B_{j , x^{[\lambda]} } ,  B_{m , x^{[\mu]} } >
< \sigma _j ^{[\lambda]}  X , \sigma _m ^{[\mu]}  Y >.$$
We have:
$$ < ({\rm d}\Gamma ( M_{\omega}) )^{-1} B_{j , x } ,  B_{m , y } >  =
\int _{\R^3} |k| ^{-1}  B_{j , x^{[\lambda]} }  (k) \cdot   B_{m , x^{[\mu]} }  (k) dk. $$
In view of definition (\ref{Bjx}), we deduce that:
\begin{align*} < ({\rm d}\Gamma ( M_{\omega}) )^{-1} B_{j , x } ,  B_{m , y } >  &=
 ( 2 \pi )^{-3} \int _{\R^3}  | \phi(|k|)|^2  e^{i   k\cdot (y - x  )}
  \frac { (k\times e_j ) \cdot (k\times e_m ) } {  |k|^2 }  dk \\
  & =  ( 2 \pi )^{-3} \int _{\R^3}  | \phi(|k|)|^2  e^{i   k\cdot (y  - x   )}
  \frac { \delta _{jm} |k|^2 - k_j k_m  } {  |k|^2 }  dk. \end{align*}
 
The proposition then follows.\hfill $\Box$

{\it Proof of Theorem \ref{th-egal}.} 
We use the Fourier transform to determine the vector-valued ${\bf B^{\rm \bf vect}}(x, X)$
from the vector-valued current density ${\bf j^{\rm \bf vect}}(x, X)$. Maxwell equations then read as:
$$   {\bf \widehat B^{\rm \bf vect}}  (\xi, X) = i \frac {\xi \times   {\bf \widehat j^{\rm \bf vect}} (\xi, X) }
{|\xi|^2}. $$
The vector field in  (\ref{dens-cour}) is divergence free.
Thus, $\xi \cdot    {\bf \widehat j^{\rm \bf vect}}(\xi, X)  = 0$. We have:
$$  | \xi \times    {\bf \widehat j^{\rm \bf vect}} (\xi, X)  |^2  =
|\xi|^2 |   {\bf \widehat j^{\rm \bf vect}}(\xi, X) |^2
 - |  \xi  \cdot   {\bf \widehat j^{\rm \bf vect}} (\xi, X) |^2  
 = |\xi|^2  |    {\bf \widehat j^{\rm \bf vect}} (\xi, X) |^2 $$
(with the  norm of  ${\cal H}_{\rm sp} ^3$). 
We deduce that:
\be\label{eg-1}     \frac {1}  {2}   \int _{\R^3 } | {\bf B^{\rm \bf vect} } (x, X) |^2  dx =   \frac {1}  {2} (2\pi)^{-3}  \int _{\R^3 }
 | {\bf \widehat B^{\rm \bf vect}} (\xi, X)|^2 d \xi = \frac {1}  {2}   (2\pi)^{-3} \int _{\R^3 }
  \frac { |   {\bf \widehat j^{\rm \bf vect}}(\xi, X) |^2}
{|\xi|^2} d \xi.  \ee 
From the current density definition (\ref{dens-cour})(\ref{rho-phi}), we see:
\be\label{eg-2}   {\bf \widehat j^{\rm \bf vect}} (\xi, X )   = i \widehat {\rho }  (\xi) \sum _{\lambda = 1}^P
 M^{[\lambda]} e^{ i x ^{[\lambda]} \cdot \xi}\  \xi \times  {\bf \sigma } ^{[\lambda]}X
 = i  \phi (|\xi|)   \sum _{\lambda = 1}^P
 M^{[\lambda]} e^{ i x ^{[\lambda]} \cdot \xi}\  \xi \times  {\bf \sigma } ^{[\lambda]}X.\ee 
Equality (\ref{maxwell-vect})  then follows  (\ref{eg-1})(\ref{eg-2}) and Proposition \ref{expl-AM}. \hfill $\Box$

{\it Proof of Theorem \ref{class-deug}}.  If  $X$ and $S$ are related by (\ref{tenso})(\ref{tenso-2}) then:
$$  <  \sigma _j^{[\lambda]}  \sigma _m^{[\mu]}  X , X >  +
 <  \sigma _m^{[\mu]}   \sigma _j^{[\lambda]}  X , X > =
\left \{ \begin{matrix}2  S_j^{[\lambda]} S_m^{[\mu]} &{\rm if}& \lambda \not= \mu \\
2 \delta _{jm} &{\rm if}& \lambda = \mu  \end{matrix} \right .. $$
Thus, in view of Proposition \ref{expl-AM}:
\begin{align}  \label{calcul}    <  A_M   X , X >  &=
 - \frac {1} {2}  \sum _{ \lambda \not= \mu }
 \sum _{1 \leq j, m \leq 3}  M^{[\lambda]}    M^{[\mu]}
 A_{jm} (x^{[\mu]}  - x^{[\lambda]}  )   S_j^{[\lambda]} S_m^{[\mu]}  -  \frac {3} {2} A_{11}(0) \sum _{ \lambda = 1 }^P (M^{[\lambda]} )^2
 \nonumber\\
&= - \frac {1} {2}  \sum _{1 \leq   \lambda , \mu \leq P }
 \sum _{1 \leq j, m \leq 3}  M^{[\lambda]}    M^{[\mu]}
 A_{jm} (x^{[\mu]}  - x^{[\lambda]}  )   S_j^{[\lambda]} S_m^{[\mu]} -   A_{11}(0) \sum _{ \lambda = 1 }^P (M^{[\lambda]} )^2.
 \end{align}
Besides, as above:
$$  \frac {1}  {2}   \int _{\R^3 }
|   {\bf B} ^{\rm \bf class} (x , S)|^2 d x  = \frac {1}  {2}  (2\pi)^{-3}  \int _{\R^3 }
| \widehat { {\bf B} }^{\rm \bf class} (\xi, S)|^2 d \xi
= \frac {1}  {2}  (2\pi)^{-3}   \int _{\R^3 } \frac { |\widehat { {\bf j} }^{\rm \bf class} (\xi, S) |^2}
{|\xi|^2} d \xi. $$
We also have:
$$   \widehat {\bf j} ^{\rm \bf class} (\xi , S) = i \phi (\xi) \sum _{\lambda = 1}^P
e^{ i \xi \cdot  x^{[\lambda]}} \xi
\times  (M ^{[\lambda]} S_{\lambda} ).  $$
Consequently:
$$ \frac {1} {2}  \sum _{1 \leq   \lambda , \mu \leq P }
 \sum _{1 \leq j, m \leq 3}  M^{[\lambda]}    M^{[\mu]}
 A_{jm} (x^{[\mu]}  - x^{[\lambda]}  )   S_j^{[\lambda]} S_m^{[\mu]}  = 
 \frac {1}  {2}   \int _{\R^3 }
|   {\bf B} ^{\rm \bf class} (x , S)|^2 d x.  $$
The proof is completed. \hfill$\Box$

As mentioned in Section \ref{s2}, we give for the reader convenience a proof of the following result which is used in the proof of Theorem \ref{class-deug-s}.

\begin{theo}\label{Hopf-S2} Fix a half-integer $s$. Let     $\Omega$ and $\pi $ be respectively given by (\ref{Omega-def}) and (\ref{Hopf-s}). Then,  $\pi $  maps  $\Omega$  into the unit sphere of $\R^3$.

\end{theo}

{\it Proof.}  We use the notations introduced at the end of Section \ref{s2}. We also use the  standard notations  $Ad, Ad^{\star}$ and $ {\cal G}^{\star} $. For any $X\in V_s$, we define a linear form $\ell _X$ on ${\cal G}$ by:
$$ \ell _X (z) =  {i} < d\rho_s (z) X , X >,\quad z\in {\cal G}. $$
For every  $X\in \Omega$, $X = \rho _s (g) X_0$ with $g\in G$,
and for all $z\in {\cal G}$, we have:
\begin{align*}
\ell _X (z) &= {i} < d\rho_s (z) \rho_s (g) X_0 ,  \rho_s (g) X_0 > \\
& =   {i} <  d\rho_s ( Ad (g) z)   X_0 , X_0  >\\
& = \ell _{X_0} ( Ad (g) z).
\end{align*}
Equivalently, $ \ell _X  = Ad^{\star}  (g) \ell _{X_0} $.
We also know that:
$$ \Vert Ad^{\star}  (g) \ell _{X_0} \Vert_{ {\cal G}^{\star} } =
  \Vert  \ell _{X_0} \Vert_{ {\cal G}^{\star} } = 1.$$
Thus, we get for any $X\in \Omega $:
 $$ \sum _{j= 1}^3 \ell _X (u_j)^2 = 1.$$
Therefore, we obtain that:
 $$   \sum _{j= 1}^3 < \sigma_j (s) X , X> ^2 = 1, $$
which proves Theorem \ref{Hopf-S2}.\hfill $\Box$

 \section{Asymptotic expansion of the ground state energy.}\label{s4}

We begin with  a classical result concerning the expectation of the number operator $N={\rm d}\Gamma(I)$ in a ground state.

\begin{prop}\label{majo-phot} There exists $C>0$ satisfying for any $|M|$ 
small enough:
\be\label{N_UM} < (  N \otimes I ) U(M) , U(M) >    \leq C |M|^2, \ee
where  $U(M)$ is a normalized ground state of $H(M)$.

\end{prop}

{\it Proof.}  One has:
$$  < (N\otimes I) U(M) , U(M) > =
\int _{\R^3} | (a(k) \otimes I) U(M) |^2 dk. $$
If $U(M)$ is a normalized ground state then one has using the Pull Through Formula (see \cite{C-G} and also \cite{SCH,G-J-71,G-polaron,G-J}):
$$ ( a(k) \otimes I) U(M) = - \frac {1} {\sqrt 2 } \sum _{\mu = 1 }^P \sum _{m = 1 }^3
 M^{[\mu]}  B _{m, x^{[\mu]}}  (k) (H(M) - E(M) + |k|) ^{-1} ( I \otimes \sigma _m^{[\mu]}) U(M), $$
 proving (\ref{N_UM}). \hfill $\Box$

We shall use a following standard inequality coming from (\ref{BFS}). There is $C>0$ 
satisfying for any $|M|\leq 1$ and any normalized ground state $U(M)$:
\be\label{majo}  \Vert (H_{\rm ph} \otimes I) U(M) \Vert \leq C |M|.\ee 

{\it Proof of Theorem  \ref{prop-2}. }  Fix $U(M)$  a normalized ground state of
$H(M)$ and  let us prove that $U(M)$ belongs to $B_{\rho |M|}  ( \Pi_{\rm sp} U(M))$ (with the notations at the beginning of  
Section \ref{s3}) with $\rho >0$ independent of $M$. Clearly: 
$$ \Vert U(M) - \Psi_0 \otimes \Pi_{\rm sp} U(M) \Vert = \Vert U(M) - \Pi_{0} U(M) \Vert
= \Vert \Pi_{\perp} U(M) \Vert \leq < ( N \otimes  I) U(M) , U(M) > ^{1/2}.$$
Then, according to Proposition \ref{majo-phot}, if $|M|\leq 1$:
$$  \Vert U(M) - \Psi_0 \otimes \Pi_{\rm sp} U(M) \Vert  \leq C |M|. $$
Besides, from (\ref{majo}):  
$$ \Vert ( H_{\rm ph}  \otimes  I) (  U(M) - \Psi_0 \otimes \Pi_{\rm sp} U(M) ) \Vert = 
\Vert  ( H_{\rm ph}  \otimes  I)   U(M) \Vert  \leq C |M|.$$
Therefore, $U(M)$ is indeed in $B_{\rho |M|} ( X_M)$
with $X_M = \Pi_{\rm sp} U(M)$ and $\rho >0$ independent of $M$. We can assume that  $\rho \geq K$ where $K$ is given by (\ref{def-K}).   
One has:
$$ \inf \sigma H(M) = \inf_{f\in D(H) \setminus \{ 0 \} } 
   \frac { < H(M) f , f >  } { \Vert f \Vert ^2 }=
\inf_{f\in B_{C|M|} ( X_M)}
   \frac { < H(M) f , f >  } { \Vert f \Vert ^2 }.     $$
If $  2C|M|  <  1$, then for any $f\in B_{C|M|} ( X)$
with $X\in {\cal H}_{\rm sp}$, we have:
 $$ \frac { 1 - 2 C  |M|  } { |X|^2 }  \leq  
 \frac { 1   } { \Vert f \Vert ^2  }  \leq 
  \frac { 1 +  8 C  |M|  } { | X|^2  }. $$
Then: 
 $$  \frac { 1 - 2 C  |M|  } { |X_M|^2 } \inf_{f\in B_{C|M|} ( X_M)}  < H(M) f , f > 
\  \leq \  \inf \sigma H(M)\  \leq \  
   \frac { 1 +8  C  |M|  } { |X_M|^2 } \inf_{f\in B_{C|M|} ( X_M)}  < H(M) f , f >. $$
Also:
 $$ |X_M|^2 + \Vert \Pi_{\perp} U(M)\Vert ^2 = 1.$$
From Proposition \ref{majo-phot}, we see that:
 $$ \Vert \Pi_{\perp} U(M)\Vert ^2\ \leq\  < (N \otimes I) U(M) , U(M) > 
\ \leq\ C |M| ^2. $$
If $ 2C|M| ^2 < 1$, we have:
 $$ 1 \leq  \frac { 1  } { |X_M|^2 } \leq 1 + 2 C  |M| ^2. $$
 Theorem  \ref{prop-2} then follows from these points together with
Proposition \ref{prop-1}. \hfill$\Box$

 \section{Proof of Theorem \ref{theo-4}.}\label{s5}

 {\it Point (i).} From Proposition \ref{majo-phot}, if $U(M)$ 
 is any normalized ground state of $H(M)$, one has:
 $$ \Vert \Pi _{\perp} U(M) \Vert ^2 \leq < ( N \otimes I) U(M) , U(M)> 
 \leq C M^2.$$
Thus, if $M_n$ goes to $0$ and if   $U(M_n)$ tends to a limit, then 
the limit is under the form $\Psi_0 \otimes X$ where $X$ has a unit norm. We have:
 $$ < A_1 X , X >  = \lim _ {n\rightarrow \infty} < A_1  \Pi_{\rm sp} U(M_n), 
  \Pi_{\rm sp} U(M_n) >. $$
We now assume that  $M^{[1]}_n = \dots = M^{[P]}_n$ which is denoted by $g_n$.
According to Proposition \ref{prop-1} and since $U(M)$ belongs to a ball $B_{\rho |M|}  ( \Pi_{\rm sp} U(M))$ as in the proof of Theorem \ref{prop-2}:
$$ g_n ^2  < A_1  \Pi_{\rm sp} ( U(M_n)), \Pi_{\rm sp} ( U(M_n))  > \  = \ < H (M_n) U (M_n) , U (M_n) > +  {\cal O}   (|M_n| ^3).$$
From Proposition \ref{prop-1} again: 
$$< H (M_n) U (M_n) , U (M_n) >  \ \leq\  g_n ^2 \inf _{|Y|= 1}  <A_1 Y , Y>  + C |M_n|^3.$$
Therefore:
$$  < A_1 X , X>  = \inf _{|Y|= 1}  < A_1 Y , Y>  $$
proving point $(i)$.

 {\it Point (ii).}  We suppose thereafter that the $P$ components of $M\in\R^P$ are equal and that the same holds for each $M_n \in\R^P$ ($n\in\N$). Suppose that the ground state eigenspace of  $H(M)$ is 
of dimension strictly greater than $k$. Then, there exist  $k+1$  normalized ground states 
  $U_j(M)$ ($j=1,\dots,k+1$) orthogonal to each other. From Proposition \ref{majo-phot}, $\Pi_{\perp}  U_j(M)$  
 is converging (in norm) to $0$ as $|M|$ goes to $0$, for each $j=1,\dots,k+1$. Thus, the norms of the  $\Pi_{\rm sp}  U_j(M)$ tends to $1$. 
Then, there is a sequence $(M_n)$ such that  $\Pi_{\rm sp}  U_j(M_n)$ has a limit  
  $X_j$ with norm $1$ as $n$ goes to infinity. Therefore,   $U_j(M_n)$ 
 tends to  $\Psi_0 \otimes X_j$. Consequently, the  $X_j$ 
  are orthogonal to each other.   In view of point $(i)$, the
  $X_j$  are eigenvectors of $A_1$ corresponding to the smallest eigenvalue of $A_1$. This leads to a contradiction if the multiplicity of this eigenvalues is less than or equal to $k$.\hfill $\Box$

        laurent.amour@univ-reims.fr\newline
Laboratoire de Math\'ematiques de Reims UMR CNRS 9008,\\ Universit\'e de Reims Champagne-Ardenne
 Moulin de la Housse, BP 1039,
 51687 REIMS Cedex 2, France.

jean.nourrigat@univ-reims.fr\newline
Laboratoire de Math\'ematiques de Reims UMR CNRS 9008,\\ Universit\'e de Reims Champagne-Ardenne
 Moulin de la Housse, BP 1039,
 51687 REIMS Cedex 2, France.

 \end{document}